\documentclass[journal=jpccck,manuscript=article, layout=twocolumn]{achemso}

\usepackage[version=3]{mhchem} 
\usepackage[T1]{fontenc}       
\usepackage[utf8x]{inputenc}
\usepackage{microtype}
\usepackage{graphicx}
\usepackage{amsmath}
\usepackage{sistyle}\SIobeyboldtrue
\usepackage{braket}
\usepackage{accents}
\usepackage{commath}
\usepackage{multirow}
\usepackage{threeparttable}

\newcommand{\vect}[1]{\accentset{\rightharpoonup}{#1}}

\author{Sébastien A. Lamarre}
\affiliation{Département de chimie, Université Laval, Québec, G1V 0A6, Canada}
\alsoaffiliation{Centre d'optique, photonique et laser (COPL)}
\author{Étienne Rochette}
\affiliation{Département de chimie, Université Laval, Québec, G1V 0A6, Canada}
\alsoaffiliation{Centre d'optique, photonique et laser (COPL)}
\author{Samuel Tremblay}
\affiliation{Département de physique, de génie physique et d'optique, Université Laval, Québec, G1V 0A6, Canada}
\alsoaffiliation{Centre d'optique, photonique et laser (COPL)}
\author{Claudine Nì. Allen}
\affiliation{Département de physique, de génie physique et d'optique, Université Laval, Québec, G1V 0A6, Canada}
\alsoaffiliation{Centre d'optique, photonique et laser (COPL)}
\email{claudine.allen@phy.ulaval.ca}

\title{CdSe Colloidal Quantum Rings}

\begin{document}

\begin{abstract}
Atomically flat semiconductor nanostructures have sharp photoluminescence emission, short radiative lifetimes and a well-defined planar structure.
However, these nanostructures lack the optical and electronic fine-tuning that justify extensive research in colloidal quantum dots. 
Here, we propose a heterostructure based on CdS nanoplatelets with laterally grown CdSe.
Samples of two different thickness and varying ring width are synthesized by a one-pot method during which CdSe ring width is controlled by Se precursor amount.
The CdSe dimensionality is continuously tuned from a 0D array of CdSe dots to a 1D CdSe quantum ring and finally a 2D CdSe nanoplatelet.
Sample characterization shows that their optical properties are tunable by both structure thickness and CdSe ring width. 
This continuous tunability extends the emission range of nanoplatelet-based structures  from \SI{428}{nm} to \SI{512}{nm} for 3 and 4 monolayers structures.

\end{abstract}

\section{Introduction}
\paragraph*{}Mesoscale rings were the object of numerous experiments exploring quantum mechanical phenomena involving phase shifts and coherence of electronic wavefunctions. 
Among others, persistent currents \cite{gunther1969flux,buttiker1983josephson}, one-dimensional weak localization \cite{shea2000electrical} and the Aharonov-Bohm effect \cite{keyser2002aharonov} were studied. 
These rings have been fabricated using various materials: superconductors \cite{doll1961experimental}, metals \cite{washburn1986aharonov}, semiconductors \cite{granados2003ga}, two-dimensional free electron gas systems \cite{fuhrer2001energy}, carbon nanotubes \cite{shea2000electrical} and graphene \cite{potasz2011electronic}. 
As an example application for these structures, Suarez \textit{et al.} used epitaxial quantum rings (QRs) as gain medium for lasers \cite{suarez2004laser}.
Until recently, these structures were not available in colloidal form. 
Since 2008, freestanding quantum wells called nanoplatelets (NPLs), nanodisks or nanobelts are accessible using colloidal synthesis, enabling a larger and cheaper source of quantum wells. 
Many colloidal II-VI semiconductor nanoplatelets having a well-defined modulable thickness and atomically flat surface have been directly prepared by solution phase methods, \textit{e.g.} CdS \cite{li2012uniform, li2009cds, ithurria2011colloidal}, CdSe \cite{ithurria2008quasi, yu2009single, ouyang2008multiple, ithurria2011colloidal, li2011size}, \ce{CdS_xSe_{1-x}} \cite{fan2015colloidal}, CdTe \cite{wang2013non,wang2009single, ithurria2011colloidal,pedetti2013optimized} and ZnS \cite{buffard2015zns}. 
Cation exchange methods also allowed the preparation of ZnSe and PbSe NPLs\cite{bouet2014synthesis}.

\paragraph*{} As the NPL growth originates from the direct reaction of precursor at the NPL periphery\cite{ithurria2011continuous}, it is possible to grow another semiconductor laterally around a NPL acting as a seed. 
One such heterostructure, CdSe/CdS core/crown, has been synthesized by Prudnikau \textit{et al.} \cite{prudnikau2013cdse} and Tessier \textit{et al.} \cite{tessier2013efficient}. 
Also, CdSe/CdTe core/crown have been prepared by Pedetti \textit{et al.} \cite{pedetti2014type} and Kelestemur \textit{et al.} \cite{kelestemur2015type}.
These former structures show good confinement because of rapid exciton transfer from the CdS crown to the CdSe core.
However, their emission is mostly dependent on the core dimensions and is not easily modulated \cite{tessier2013efficient}.
To extend the emission window of NPL based heterostructures, we propose CdSe colloidal QRs, thereafter cQRs, grown at the periphery of CdS NPLs. 
As expected from the band alignment, the emission of these heterostructures originates from the excitonic recombination in the CdSe QR. 
Such colloidal quantum rings have recently been reported in parallel by Delikanli \textit{et al.} \cite{delikanli2015continuously} but prepared by a different method and only with 4 monolayers (ML) thick CdS seed.
In comparison, the presented method is a user-friendly one-pot synthesis and was used to prepared both 3 and 4 ML heterostructures. 
Colloidal QRs, as their core/crown analog, could be used in light harvesting and charge separator for solar cells, active medium for laser \cite{guzelturk2014amplified} and light-emitting diode \cite{vashchenko2014organic}, and even luminescent probe for biomedical imaging. 
Furthermore, the peculiar electronic structure of QRs has an importance for more fundamental physical studies \cite{orellana2003conductance, bayer2003optical, keyser2002aharonov, fuhrer2001energy} and light matter interaction \cite{sen2007electron, warburton2002giant}.

\paragraph*{}Here, we present the synthesis and characterization of CdSe colloidal QRs grown around 3 and 4 monolayers. 
The samples are characterized by elemental analysis, X-ray diffraction, transmission electron microscopy, absorption, photoluminescence emission and photoluminescence excitation spectroscopy. 
We describe the evolution of the CdSe dimensionality with the width of the CdSe QR from 0D to 2D with an intermediate 1D state.

\section{Experimental Methods}
\paragraph*{Chemicals}Cadmium acetate dihydrate \\(\ce{Cd(Ac)2*2H2O}, \SI{98}{\%}), technical grade 1-octadecene  (ODE, \SI{90}{\%}) and technical grade tri-$n$-butylphosphine (TBP, \SI{95}{\%}) were purchased from Alfa Aesar. 
Technical grade oleic acid (OA, \SI{90}{\%}) and elemental selenium (\SI{99.5}{\%}) were purchased from Sigma Aldrich. 
Chloroform and methanol were purchased from BDH and elemental sulfur (\SI{99}{\%}) from Laboratoire Mat.
All chemicals were used as purchased without further purification. 

\paragraph*{Preparation of stock solutions} Sulfur stock solution was prepared by dissolving \SI{32}{mg} of elemental sulfur in \SI{20}{g} of ODE by sonication at room temperature.
\SI{0.1}{M} Se stock solution was prepared by heating, under inert atmosphere, \SI{318}{mg} of elemental selenium in \SI{40}{mL} of ODE at \SI{180}{\degC} overnight. 
The resulting solution was yellow and was conserved under ambient conditions.

\paragraph*{\SI{4}{ML} CdS nanoplatelet synthesis} In a \SI{250}{mL} reaction flask, \SI{2}{mmol} of cadmium acetate, \SI{10}{g} of S stock solution, \SI{2}{mmol} of OA and 30 g of ODE were degassed under vacuum at \SI{50}{\degC} for \SI{30}{min} and purged twice with nitrogen. 
The mixture was then heated under nitrogen to \SI{260}{\degC} in about \SI{15}{min}. 
The mixture reacted at this temperature, but not above, for \SI{1}{min} before being cooled down to ambient temperature. 
At the beginning of the cooldown, \SI{3}{mL} of OA was added to improve the purification process and the stability of the purified nanoplatelets. 
After the cooldown, the mixture was centrifuged at \SI{6000}{rcf} for \SI{6}{min} and the precipitated nanoplatelets were redispersed in \SI{20}{mL} of chloroform.

\paragraph*{\SI{3}{ML} CdS nanoplatelet synthesis}\SI{3}{ML} CdS nanoplatelets were prepared in a similar manner.
Instead of oleic acid, the same molar amount of myristic acid was used for the reaction mixture.
For this synthesis, the reaction time and temperature were \SI{20}{min} and \SI{180}{\degC}.

\paragraph*{CdSe ring synthesis around CdS nanoplatelets} In a typical synthesis for cQRs, \SI{0.1}{mmol} of cadmium acetate, \SI{15}{mg} of oleic acid, \SI{0.5}{mL} of a CdS NPL dispersion, \SI{0.5}{mL} of the Se stock solution and \SI{15}{mL} of ODE were added to a \SI{100}{mL} reaction flask, degassed at \SI{50}{\degC} and purged twice with nitrogen. 
The mixture was heated to \SI{190}{\degC} and kept at this temperature, never above, for \SI{10}{min}. 
Afterward, \SI{1}{mL} of oleic acid was injected and the mixture was cooled to room temperature. The mixture was centrifuged at \SI{11000}{rcf} for \SI{8}{min}. 
The precipitated cQRs were redispersed in \SI{6}{mL} of chloroform. The dispersions are colloidally stable for months.

\paragraph*{}The ring width is controlled by the amount of Se added. 
This amount can be lowered down to \SI{5}{\micro mol} to obtain a narrower ring. 
Above \SI{50}{\micro mol}, the concentration of selenium was too high and secondary CdSe NPL and dot nucleation hindered ring growth.
To obtain the sample with more than \SI{50}{\micro mol} of Se, cQRs with \SI{50}{\micro mol} of Se were prepared according to the method.
After the \SI{10}{min} reaction time, the remaining Se quantity, up to an additional \SI{50}{\micro mol} was injected in the reaction flask and the temperature was kept at \SI{190}{\degC} for 10 more minutes. Multiples injections were used  for samples above \SI{100}{\micro mol} of Se.

\paragraph*{Energy-dispersive X-ray spectroscopy (EDS)}EDS measurements were obtained with a JEOL JSM-840-A scanning electron microscope using a PGT Avalon EDS with a NORA detecting unit. 
The samples were scanned with a \SI{15}{kV} electron beam. 

\paragraph*{Powder X-ray Diffraction (XRD)}Diffractograms were acquired using a Siemens-Bruker X-ray diffractometer with a 2D Hi-Star XRD detector. 
The radiation source was a Kristalloflex 760 with a nickel window emitting the Cu K$_{\alpha}$ line ($\lambda = \SI{1.5418}{\angstrom}$) with an accelerating voltage and current of respectively \SI{40}{kV} and \SI{40}{mA}.
Diffractograms were recorded from \ang{10} to \ang{60}. 
Background signal was automatically subtracted by the diffraction pattern treatment software GADDS.

\paragraph*{Purification method for EDS and XRD} cQR dispersions were centrifuged at low force to eliminate the remaining insoluble compounds. 
Afterward, the cQRs were precipitated in a solution of \SI{10}{\%} TBP in methanol and centrifuged. 
The precipitate was dispersed with chloroform. 
This precipitation cycle was repeated once more with the TBP solution and once again with pure methanol as precipitating solvent. 
Finally, the cQRs were redispersed in a minimal amount of chloroform and drop-casted on a Si wafer for the EDS and on a glass cover slip for XRD analysis.

\paragraph*{Transmission electron microscope (TEM)}The purified cQR samples morphology was characterized with a JEOL 1230.
The samples dispersed in chloroform were drop-casted on a Ni TEM grid coated with Formvar and carbon film.

\paragraph*{Optical spectroscopy}All optical characterizations were carried out on cQR dispersions in chloroform. 
Absorption spectra of the samples were recorded at room temperature with a Varian Cary 50 Conc UV-visible spectrophometer from 300 to \SI{800}{nm}.
The steady state photoluminescence (PL) and photoluminescence excitation (PLE) spectra were acquired using a Jobin-Yvon Fluorolog equipped with a photomultiplicator tube. 
The samples were excited at \SI{350}{nm} for PL spectra and the detection window was set at the emission maximum for PLE spectra. 
Excitation and emission monochromators slits were set at \SI{1}{nm}.

\paragraph*{Photoluminescence peak fitting} PL spectra were fitted to extract their respective photoluminescence maximum energy (PL$_\text{max}$) and full width half maximum (FWHM).
Depending on the Se amount added during the cQR synthesis, the PL behaviour might correspond to either a Gaussian or Lorentzian peak shape, which correspond to different broadening types, as discussed in Spectroscopic Characterization section.
Thus, all PL spectra were fitted with both peak shapes on an interval corresponding to the exitonic peak.
Then, for each spectrum, the most representative peak shape was determined by comparing each fit's coefficient of determination ($R^2$) over the whole spectrum.
From this best fit, the PL$_\text{max}$ was directly obtained from the fit's center parameter and the FHWM was calculated from the fit's width parameter.

\section{Results and Discussion}

\subsection*{Synthesis and Expected Growth Mechanism}

\paragraph*{}The synthesis of these CdSe cQRs involves two steps.
First, CdS nanoplatelets of \SI{3}{ML} and \SI{4}{ML} are prepared following a one-pot method reported by M. Li \textit{et al.} \cite{li2009cds} and Z. Li \textit{et al.} \cite{li2012uniform}.
Compared to continuous injection methods, this approach is more easily implemented for various experimental conditions and facilitates synthesis scale up.
Second, the growth of a CdSe ring around the CdS NPL periphery is also inspired from a CdSe NPL synthesis \cite{li2011size}.
The one-pot method presented has the advantage of being more user-friendly than continuous injection method and yields reproducible results as all the results were replicated by another user.
As with CdSe NPL synthesis, the reaction conditions are critical for the growth of atomically flat CdSe. 
The range of synthetic temperatures is below the solubilization temperature of the CdS and CdSe NPLs and higher than the monomer reaction temperature/sout{. The reaction temperature has to be higher than the monomer reaction temperature} (\SI{140}{\degC})\cite{li2011size}.
At higher temperatures (\SI{240}{\degC}), the NPLs are not stable and the formation of irregular free CdSe nanocrystals is more favorable than lateral growth.\cite{li2011size}
The Se precursor concentration is kept low to promote heterogeneous nucleation and growth of CdSe around CdS NPLs over secondary homogeneous CdSe nucleation.

\begin{figure}[!htbp]
   \centering
    \includegraphics[width=0.8\columnwidth]{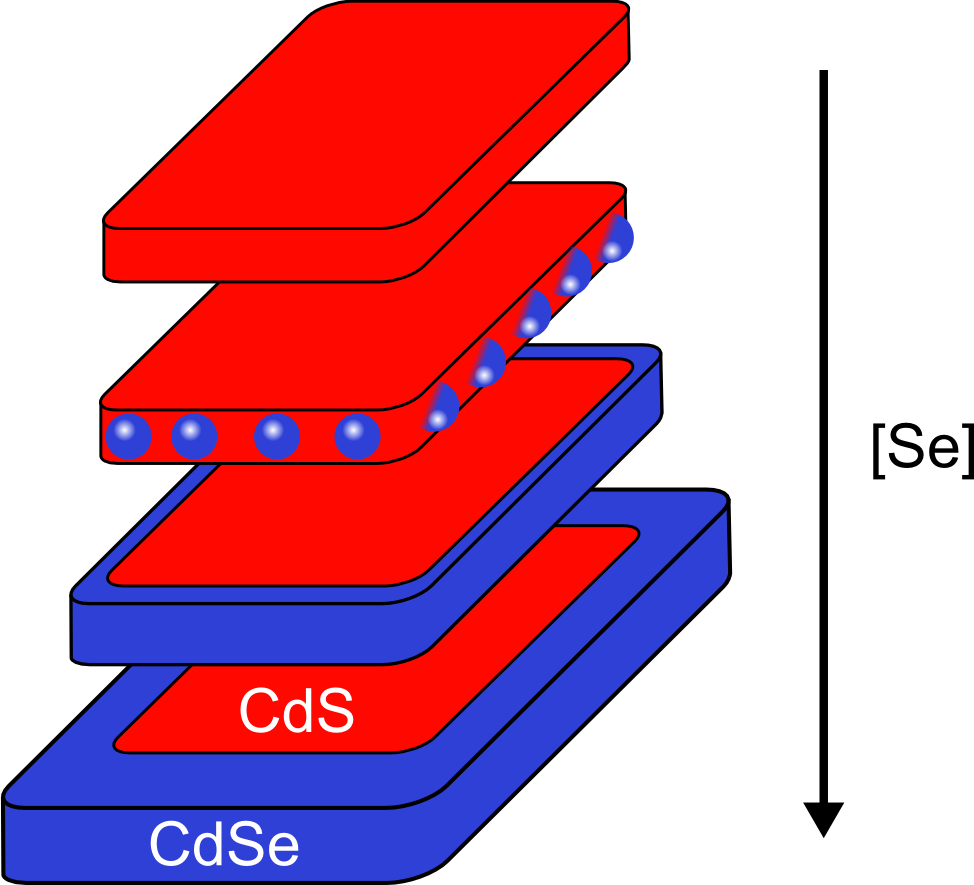}
          \caption{Evolution of the cQR shape as function of the Se amount present in the synthesis.}
\end{figure}

The oleic acid concentration is also crucial to grow NPLs laterally instead of wires or dots \cite{srivastava2010surface}.  
Indeed, the synthetic conditions with saturated \ce{Cd(Ac)2} combined with a suitable concentration of free oleic acid promotes lateral growth of CdSe. 

\paragraph*{} Both the reaction conditions and the observed CdSe ring growth suggest the reaction mechanism to be analogous to CdSe NPL growth as described by Ithurria \textit{et al.} \cite{ithurria2011continuous}: monomers react with the CdS or CdSe peripheral edges to grow even further. 
The monomers can be either cadmium carboxylate salts and dissolved Se or small building blocks containing both Cd and Se as proposed by Yu \textit{et al.}\cite{yu2010thermodynamic,yu2012cdse}  for CdSe magic-sized clusters. 
The lateral extent of the cQRs is controlled by the amount of Se, because the growth mechanism is the constant reaction of monomers and diffusion controlled.
When there is not enough CdSe to cover the entire periphery of every CdS NPL, CdSe forms dot-like structures on the edges of the CdS NPLs as illustrated in the top part of Fig. 1 and proved in the section Spectroscopic Characterization.
Following epitaxial growth terminology, these dot-like structures are named islands.
Assuming the CdSe heterogeneous nucleation is evenly distributed around every NPL, at first the growth is likely to proceed locally without a wetting layer (Volmer-Weber growth mode).  
With sufficient amount of Se precursor, each island can grow in the direction of the width or extend itself around the periphery of the CdS NPL to form a complete ring. 
Also, this competition between the two growth directions is apparent spectroscopically with a larger PL FWHM than an uniform and flat CdSe layer. 
Once the islands have connected with each other as represented in the bottom part of Fig. 1, the growth is limited to the lateral direction and the CdSe grows like typical NPLs.

\subsection*{Material Characterization}
\paragraph*{}Elemental and structural analyses are performed on the purified samples to corroborate that CdSe grows around the rim of the CdS NPLs. 
Table 1 presents the energy-dispersive X-ray spectroscopy (EDS) results that confirm a higher Se fraction is found in wider cQRs grown from a larger initial quantity of Se precursors. 
There is an excess of Cd even when considering both basal planes are metal-terminated: the Cd molar fraction should be \SI{57}{\%} for \SI{3}{ML} cQRs, whereas the measured amount stated in Table 1 is around \SI{70}{\%}. 

\begin{table*}
\centering 
\begin{threeparttable}
\caption{Photoluminescence maximum, elemental analysis and estimated ring size of selected \SI{3}{ML} cQRs}
\begin{tabular}{cccccc}
\hline
\multirow{2}{*}{Se added}& \multirow{2}{*}{PL$_\text{max}$} & \multicolumn{3}{c}{Elemental analysis} & \multirow{2}{*}{Estimated ring width\tnote{a}} \\
 & & Cd & S & Se & \\ \hline
\SI{}{\micro mol} & eV (nm)  & \% & \% & \% & nm \\ \hline
10 & 2.879 (430.6)& 72 & 27 & 1 & 0.3 \\
50 & 2.703 (458.7)& 71 & 20 & 9 & 2.9 \\
100 & 2.692 (460.6)& 68 & 14 & 18 & 7.9 \\ \hline
\end{tabular}
\begin{tablenotes}
	\item [a] Considering \SI{30}{nm} wide NPLs with square geometry and a complete covering.
\end{tablenotes}
\end{threeparttable}
\end{table*}

\paragraph*{}This excess beyond the expected stoichiometry likely originates from Cd organic salts left in solution after the purification procedure. 
As opposed to the excess Cd fraction, the Se fraction cannot come from Se impurities because the purification method with TBP efficiently removes the Se precursor.
Thus, the increasing Se fraction is directly linked to CdSe growth.
The ratio between S and Se molar fraction was used to estimate the CdSe ring width with the assumptions that the CdS NPLs are \SI{30}{nm} large squares and that CdSe forms an uniform layer.
These calculations could not be confirmed because of the lack of contrast between CdSe and CdS in TEM micrographs, thus the lack of direct CdSe ring width measurements.
Overall, the elemental analysis indicates that cQRs comprise  both CdS and CdSe in agreement with the XRD pattern shown in Fig. S3\dag.
The XRD pattern matches the lines in the reference patterns of zinc blende CdS and zinc blende CdSe with no feature corresponding to a wurtzite phase.
The zinc blende crystalline phase is typical of Cd chalcogenides grown in  NPL structures \cite{li2011size}. 
The corresponding atomically flat morphology is confirmed by the TEM image of Fig. 2b where a single population of NPL-like structures is observed without any apparent secondary nucleation. 

\begin{figure}[!htb]
   \centering
    \includegraphics[width=0.9\columnwidth]{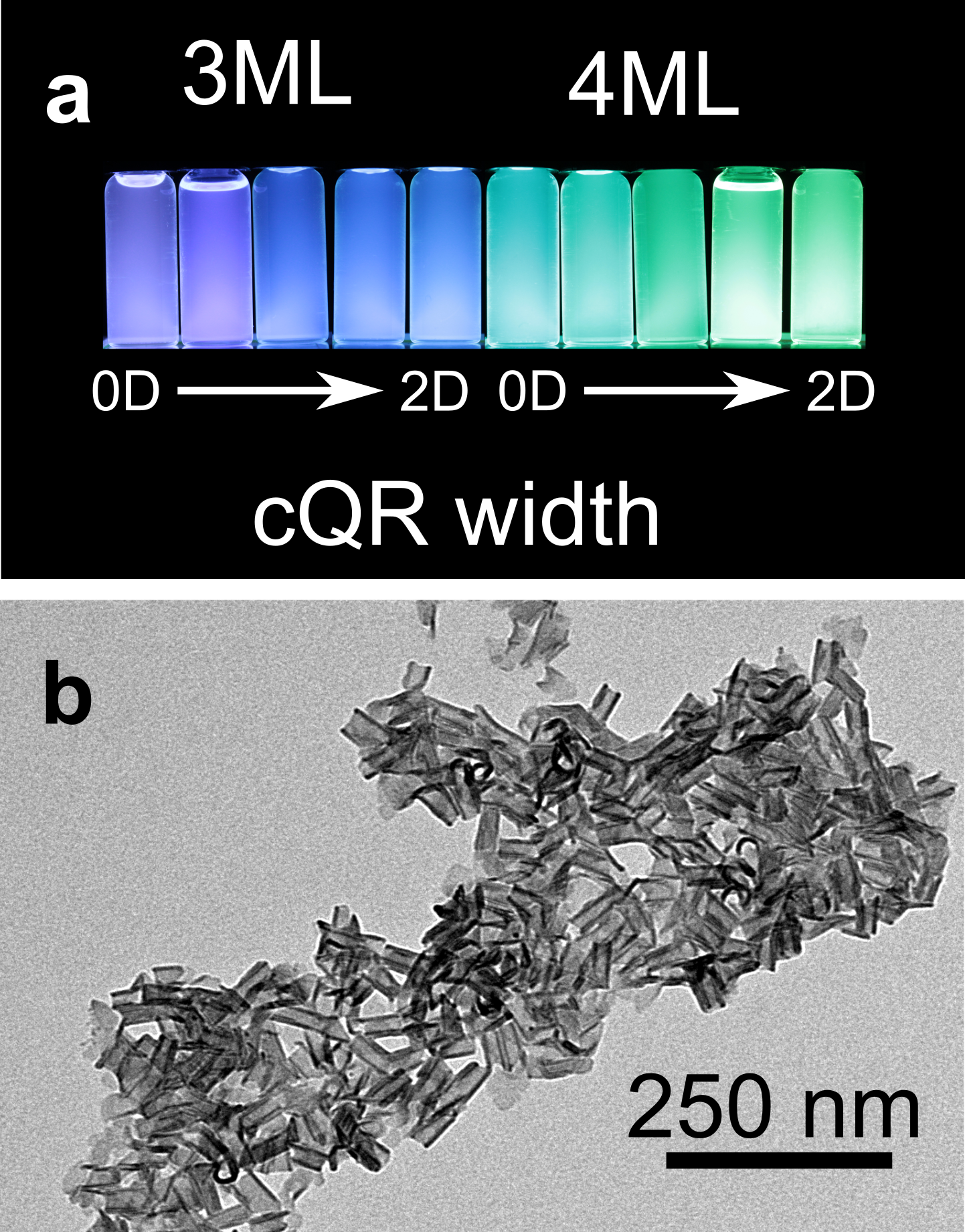}
          \caption{(a) Photographs of \SI {3}{ML} and \SI {4}{ML} cQRs under UV excitation. Ring width increases from left to right. (b) Transmission electron microscopy image of \SI {4}{ML} cQRs with \SI{50}{\micro mol} of Se.}
\end{figure}

\paragraph*{}These results and previous reports on lateral heterostructures grown in similar reaction conditions \cite{prudnikau2013cdse, tessier2013efficient, delikanli2015continuously} point to a morphology where CdSe surrounds the CdS NPLs in the plane, which is also supported by the spectroscopic characterization (section below).
\subsection*{Spectroscopic Characterization}

\begin{figure}[!htb]
   \centering
    \includegraphics[width=\columnwidth]{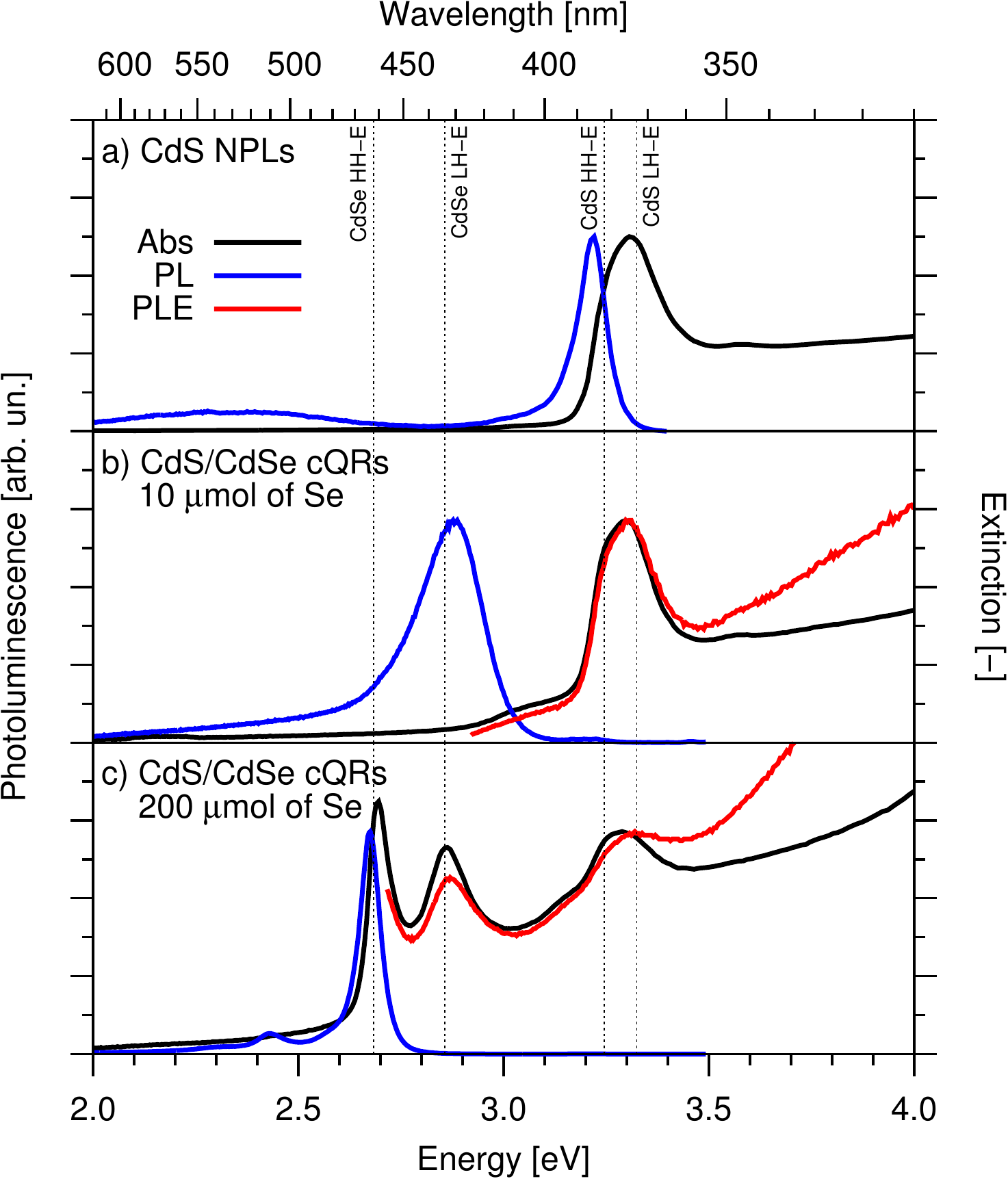}
          \caption{Absorption, photoluminescence and photoluminescence excitation spectra of (a) \SI {3}{ML} CdS nanoplatelets, (b) \SI {3}{ML} cQRs with \SI{10}{\micro mol} of Se and (c) \SI {3}{ML} cQRs with \SI{200}{\micro mol} of Se. HH-E and LH-E transitions for \SI {3}{ML} CdS and CdSe NPLs\cite{tessier2013efficient} are identified with dotted lines.}
\end{figure}

\paragraph*{}Further investigations with optical spectroscopy confirm that an electronically conjugated planar heterostructure is indeed synthesized. 
In Fig. 3a, the absorption and PL spectra excited at \SI{350}{nm} are presented for \SI{3}{ML} CdS NPLs used as seeds in the CdSe ring growth process. 
The electron-heavy hole (HH-E) and electron-light hole (LH-E) transition energies for \SI{3}{ML} CdS NPLs\cite{tessier2013efficient}, respectively at \SI{3.246}{eV} (\SI{382.0}{nm}) and \SI{3.324}{eV} (\SI{373.0}{nm}), are indicated by dashed lines and correspond to the overlapped peaks in the absorption spectrum. 

\paragraph*{}The spectral characteristics of the cQR samples in Fig. 3b,c are notably different from those of \SI{3}{ML} CdS NPLs with the emergence of absorption and emission at energies below the CdS NPL spectra.
These features are attributed to the CdSe added around the NPLs. 
The CdS NPL emission is actually quenched even without selective excitation of the CdSe rings and with a minimal amount of CdSe.
As soon as the CdSe nucleates into islands on the NPL rim with \SI{10}{\micro mol} of Se precursor added, the emission peak shifts from \SI{3.220}{eV} (\SI{385.0}{nm}) to \SI{2.861}{eV} (\SI{433.4}{nm}) with a final FWHM of \SI{215}{meV} (\SI{32.6}{nm}) between Fig. 3a and Fig. 3b. 
When the Se amount reaches \SI{200}{\micro mol} in Fig. 3c, the PL emission peak has redshifted down to the energy of the HH-E transition of CdSe NPLs at \SI{2.67}{eV} (\SI{464}{nm}), with a FWHM of \SI{57}{meV} (\SI{9.8}{nm}), also typical of atomically flat 2D CdSe NPLs. 
The low energy emission tail originating from trap states in islands with a large surface-to-volume ratio has also disappeared. 
An extra peak is then observed at \SI{2.43}{eV} (\SI{510}{nm}) and is attributed to thicker \SI{4}{ML} CdSe 2D nanostructures either grown around thicker CdS secondary nucleation or direct CdSe structure nucleation.
For \SI{200}{\micro mol} sample, the absorption spectrum now features LH-E (\SI{2.863}{eV}, \SI{433.1}{nm}) and HH-E (\SI{2.695}{eV}, \SI{460.1}{eV}) transitions corresponding to isolated \SI{3}{ML} CdSe NPLs and the combined LH-E and HH-E transitions (\SI{3.29}{eV}, \SI{377}{nm}) of \SI{3}{ML} CdS NPLs as labeled in Fig. 3 \cite{tessier2013efficient}.
The photoluminescence excitation spectrum with a detection window centered on the PL maximum is also provided in each panel to show its close match to the absorption spectrum from the CdSe band edge up to \SI{3.5}{eV} (\SI{350}{nm}). 

\begin{figure*}[!htb]
   \centering
    \includegraphics[width=0.8\textwidth]{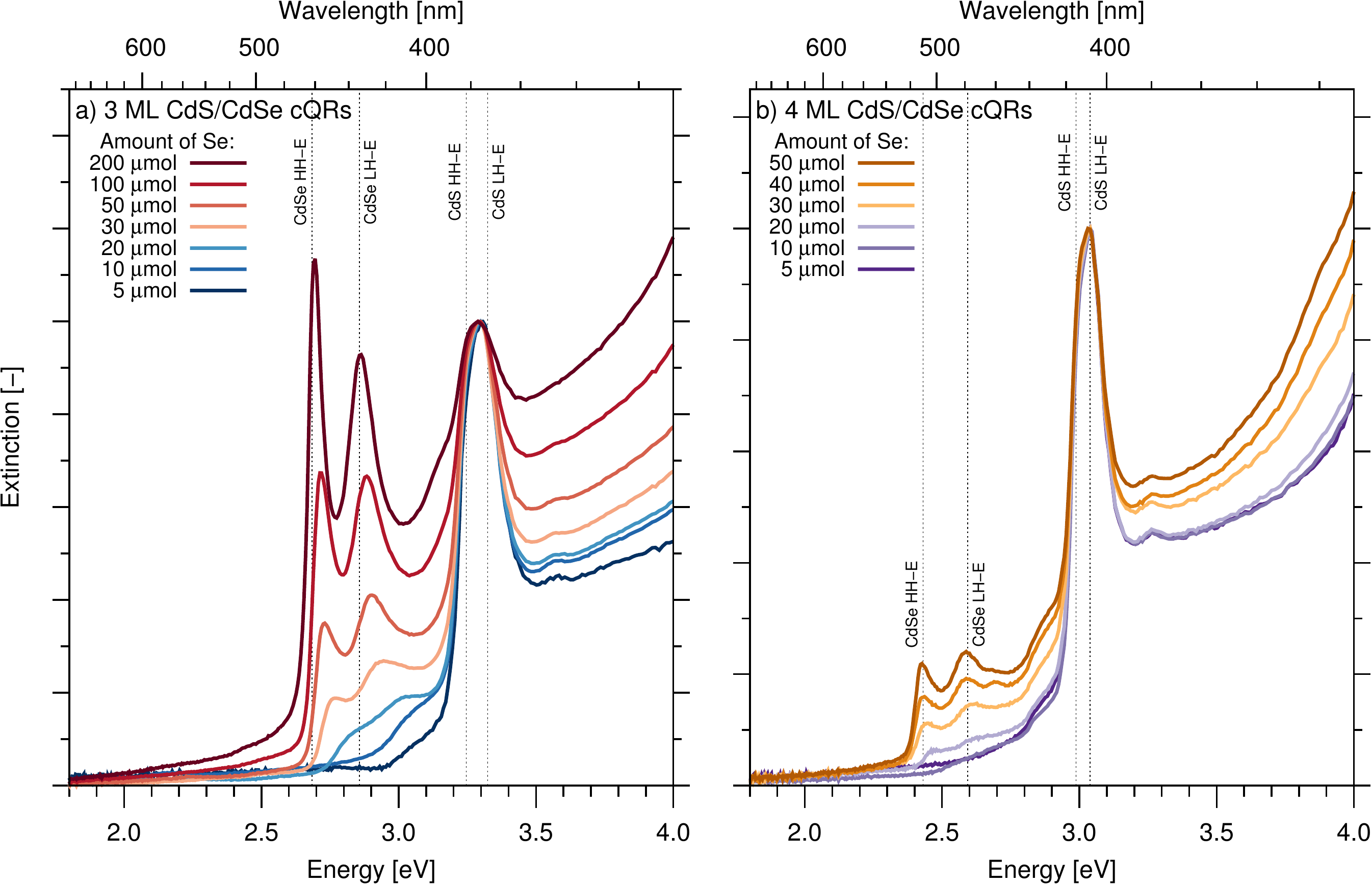}
          \caption{Absorption spectra of (a) \SI {3}{ML} and (b) \SI {4}{ML} cQR samples. The spectra are normalized to the CdS NPL absorption transition.}
\end{figure*}

\paragraph*{} While Fig. 3b,c give an overview of the spectral characterizations done at both extreme Se amounts for \SI{3}{ML} CdS/CdSe heterostructures, the next Fig. 4-6 show the detail of those measurements for all Se doses and heterostructure thicknesses.
In Fig. 4, the CdSe absorption features gradually appear as more CdSe is grown around the CdS NPLs for both \SI{3}{ML} and \SI{4}{ML} CdS NPLs.
The CdSe is more apparent for \SI{3}{ML} samples as the CdS NPL seeds used are smaller than the \SI{4}{ML} ones. 
With small Se amount, the CdSe absorption is very weak and only one absorption peak can be seen. 
Starting at \SI{20}{\micro mol}, the HH-E and LH-E transitions are more visible but they are much larger and blue-shifted than those expected for CdSe NPLs. 
At \SI{200}{\micro mol}, the transitions have shifted back to the characteristic energies of CdSe NPL transitions.
PLE spectra showing equivalent behaviour are presented in Fig. S4\dag.
PL spectra for all Se amounts and thicknesses are shown in Fig.5, in which both the PL$_\text{max}$ and FWHM shifts are apparent. 
To extract quantitative information about those shifts, the PL spectra were fitted as described in the Experimental Methods section. The best fit PL$_\text{max}$ and FWHM are shown in Fig. 6.
These quantities are correlated, for both thicknesses: as the PL$_\text{max}$ shifts to the lower energy, so does the FWHM.
The best fit for lower Se amount PL spectra was systematically the Gaussian peak shape, while the Lorentzian peak shape was more suited to the higher Se dose PL spectra. For each thickness, the PL$_\text{max}$ and FWHM shifts are reduced after the change from Gaussian to Lorentzian peak shape.

\paragraph*{} The equivalence between the absorption and PLE spectra is shown in Fig. 3 and can be inferred by comparing Fig. 4 and S2. 
This is attributed to ultrafast non-radiative relaxation bringing the photogenerated charge carriers to the lowest excited and emissive states in CdSe near the band edge where the PLE signal is recorded \cite{tonti2004excitation}, resulting in the CdS PL quenching.
This is coherent with Tessier \textit{et al.}\cite{tessier2013efficient} works with CdSe/CdS core/crown.
This demonstrates the formation of CdS/CdSe heterostructures with charge carrier diffusion between both materials instead of independent nanocrystal populations.

\paragraph*{} Comparison of the absorption and PLE normalized spectra shows that CdSe PLE peaks are less intense than their respective absorption peaks. 
This observation could be the result of either secondary nucleation of CdSe NPLs, which have a lower quantum yield than CdSe rings, or a smaller quantum yield of the CdSe rings when directly excited. 
In both cases, the diminished CdSe quantum yield would result in weaker CdSe PLE compared to CdSe absorption, whereas the CdS peaks would remain unchanged.  
Importantly, the comparison of the normalized absorption and PLE spectra indicates that no isolated CdS NPLs are present in the sample. 
Indeed, the presence of isolated CdS NPLs would increase CdS absorption while CdSe absorption of the overall sample would stay the same, thus decreasing the CdSe absorption when normalized to that of CdS. 
On the other hand, any ringless CdS NPLs would not contribute to PLE as the acquisition window only probes CdSe emission.  
The presence of CdS NPLs would therefore lead to the normalized CdSe PLE peaks being more intense than the normalized CdSe absorption, which is the opposite of what is observed.
In the latter case, the spectral signature of remaining CdS NPLs was observed
This indicates that the non-injection synthesis employed in the present study has the advantage of producing more well-defined heterostructures than the continuous injection method. 

\begin{figure}[!htb]
   \centering
    \includegraphics[width=0.9\columnwidth]{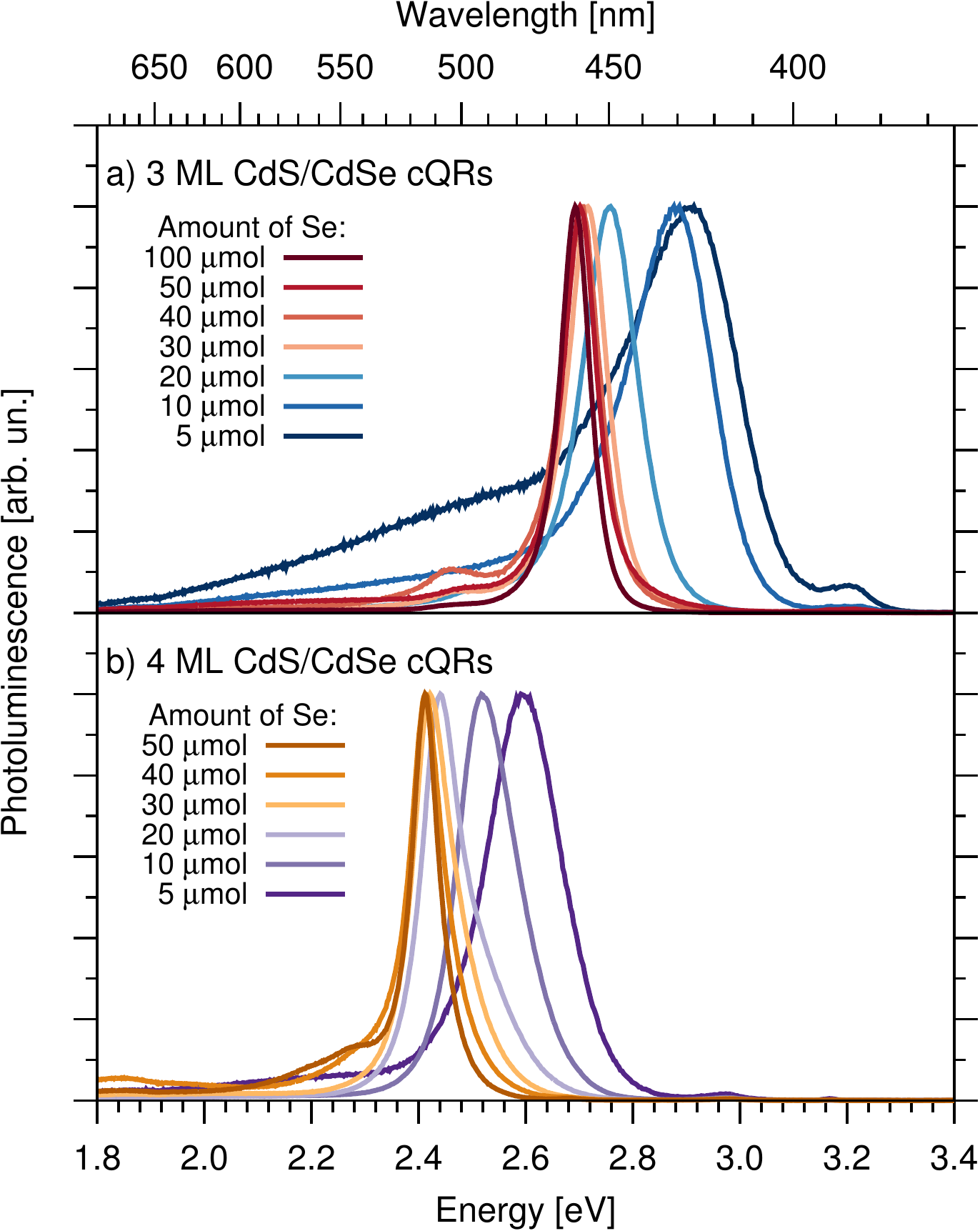}
          \caption{Photoluminescence spectra of (a) \SI {3}{ML} and (b) \SI {4}{ML} cQRs with increasing amount of Se added to the synthesis.}
\end{figure}

\paragraph*{} In addition to the proposed ring geometry, other heterostructure geometries could generate these results.
We will now demonstrate that the proposed geometry is the most likely candidate by refuting the two most probable geometry alternatives: CdS/CdSe core/shell and CdS/CdSe heterodimer.
Knowing that the CdS and CdSe are electronically connected, it is possible that a CdSe NPL has grown on top the CdS NPL instead of being connected to its rim, as in the case of core/shell CdSe/CdS NPLs\cite{mahler2012core}. 
If it was the case, a striking redshift and a wider bandwidth of the CdSe absorption and emission transitions would be observed as demonstrated by Mahler \textit{et al.} \cite{mahler2012core} for CdSe NPLs with a CdS shell. 
However, the CdSe emission is precisely the same as CdSe NPLs of the same thickness as its CdS NPL seeds without a shell\cite{mahler2012core}. 
This confirms that CdSe is laterally attached to the CdS NPLs. 
Two planar geometry limit cases are possible: the proposed CdSe rings or heterodimers\cite{de2011size}.
The ring geometry is more probable as our cQR synthesis is very analogous to the core/crown synthesis\cite{mahler2012core} and imply a similar reactivity for all facets of the periphery.
Furthermore, the quantum ring geometry was demonstrated by line-EDX for continous injection grown CdS/CdSe heterostructures by Delikanli \textit{et al.}\cite{ delikanli2015continuously}.

\paragraph*{}For larger Se amounts, the CdSe transitions have negligible inhomogeneous broadening resulting in a Lorentzian band shape and a small FWHM. 
This is observed in Fig. 5 for both \SI{3}{ML} and \SI{4}{ML}. 
The sharp absorption peaks of the LH-E and HH-E transitions are distinct and intense (Fig. 4). 
For the higher Se amount, the CdSe have the electronic structure of atomically flat CdSe NPLs.
For small Se amounts, the CdSe islands around all CdS NPLs are randomly dispersed in size, resulting in a normal distribution of energy levels. 
This inhomogeneous broadening translates in a broader Gaussian shaped band, often observed for ensemble colloidal quantum dots (cQDs), as shown in Fig. 5. 
The absorption spectra of these samples do not have the LH-E and HH-E transitions of a quantum well but instead the broad $1S_e-1S_{3/2}$ transition of cQD ensembles. 
However, the very small volume of CdSe islands compared to the CdS NPLs on which they reside makes their absorption appear only as a weak tail in Fig. 4. 
In conjunction with both thermal and inhomogeneous broadening, it is thus difficult to discern distinct transition peaks for CdSe islands.
In light of these observations, these samples form CdSe dots around CdS NPLs.

\subsection*{Overview of Synthesized Samples}

\paragraph*{}The series of lateral CdS/CdSe heterostructures prepared emit in the higher energy portion of the visible spectrum, yielding the bright purple, blue and green PL shown in Fig. 2a. 
Whereas such colors are obtainable from a number of nanocrystal types such as CdSe cQDs, optimizing their nucleation-growth to obtain smaller sizes for short-wavelength emission is more demanding.
Indeed, a large concentration of precursors is required to initiate the nucleation process, in turn calling for short growth times to keep the cQDs small. 

\paragraph*{} In the lateral heterostructure cQR case, CdSe growth onto the CdS NPL atomically thin edges is reminiscent of deposition and epitaxial growth processes on crystalline substrates. 
In this heterogeneous growth regime, the lateral extent is limited by the amount of Se precursor and without restrictions on reaction time for CdSe growth.
The resulting heterostructures are thin colloidal sheets curled into tubes as shown in Fig. 2b, just like the homogeneous NPLs \cite{bouet2013two}.
While control of the flat NPL thickness at the atomic level causes discrete jumps of the emission energy, as illustrated from blue colors  to green ones in Fig. 1a corresponding to 3 and 4 ML respectively, the continuous PL tuning is recovered through the lateral extent of CdSe grown.
Therefore, cQRs provide two degrees of freedom to control the quantum confined energy levels.
Indeed, the thickness can be set to obtain emission from CdSe near the desired color and its width controls the PL bandwidth while fine tuning the emission wavelength.

\subsection*{Wavefunction Dimensionality Evolution}

\paragraph*{}As presented in Fig. 6, the sample series behaviour changes drastically as the Se amount increases. 
This is attributed to a change in the CdSe dimensionality, \textit{i. e.} the degrees of freedom in which the charge carriers can move in the CdSe.
These varying degrees of freedom originate from a varying description of the charge carrier envelope wavefunction.
If charge carriers are free in a direction (\textit{e. g.} $x$), they are approximated as delocalized  in that direction and, as in a bulk crystal, $\Ket{\vect{k}_x}$ is an adequate description of the wavefunction in $x$.
Inversely, if the charge carriers are confined in $x$, the bound states of the particle in a box model $\Ket{n_x}$ are more suitable to describe the wavefunction in $x$.

\begin{figure}[!htb]
   \centering
    \includegraphics[width=\columnwidth]{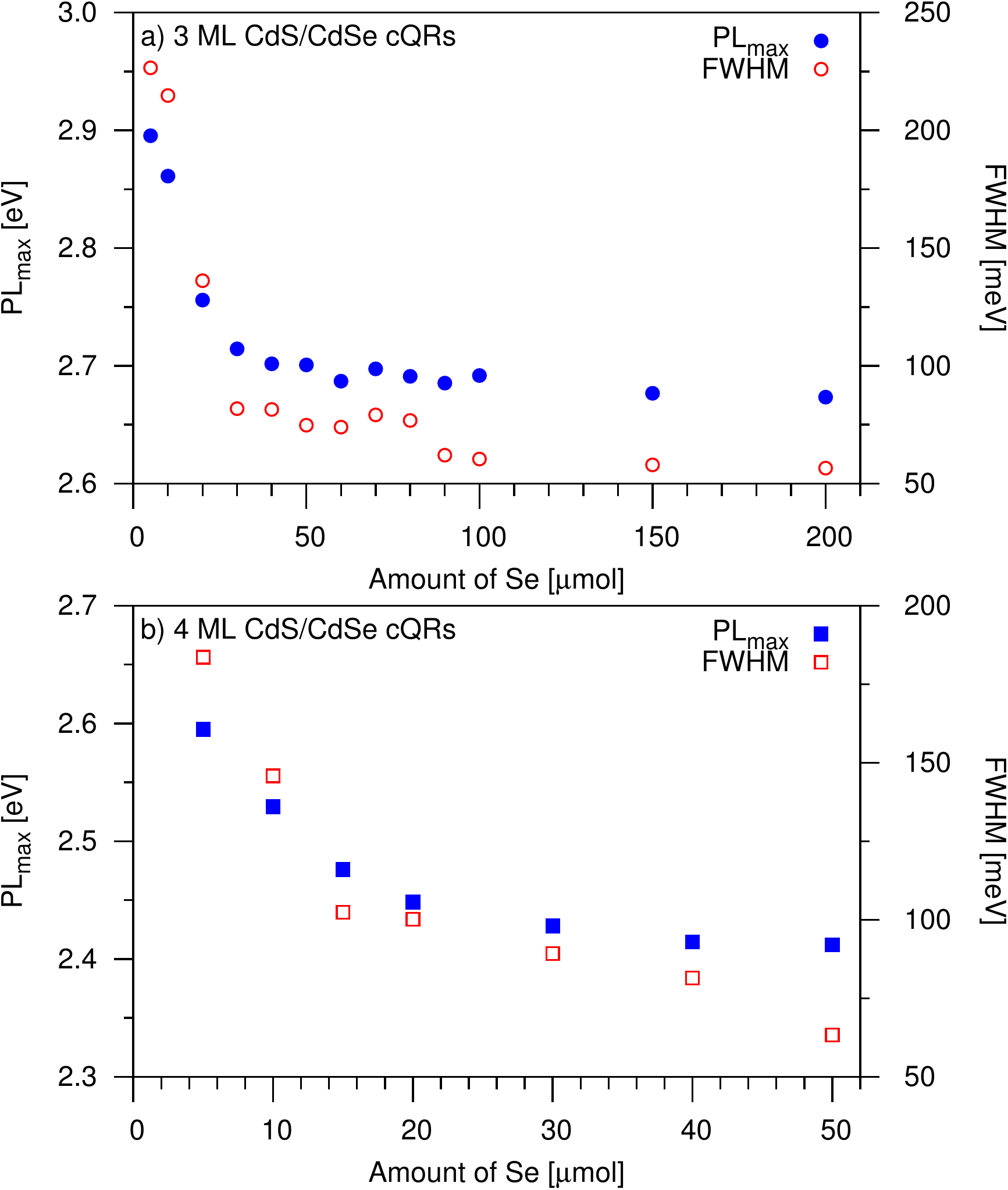}
          \caption{Energy of the photoluminescence maximum (filled symbols) and FWHM (empty symbols) of (a) \SI {3}{ML} and (b) \SI {4}{ML} cQRs. The parameters are extracted from the best fit between a Gaussian and a Lorentzian peak shape. The Gaussian peak shape is preferable at lower Se amount, specifically up to \SI{20}{\micro mol} for \SI {3}{ML} and \SI{10}{\micro mol} for \SI {4}{ML} cQRs}.
\end{figure}

\paragraph*{} In the CdSe islands resulting from a low Se amount, the charge carriers have no degrees of freedom and their state can be described by $\Ket{n_x, n_y, n_z}$. 
At high Se amount, the charge carriers are only confined in $z$. 
Thus, they have two degrees of freedom and can be described by $\Ket{\vect{k}_x,\vect{k}_y,n_z}$.
Between these two extreme cases, the system goes through a 1D configuration with periodic boundary conditions with charge carriers diffusing around the ring. 
It is this intermediate regime that is defined as the cQR.
The most appropriate representation of the envelope wavefunction in this case would be that of a cylindrical box with Bessel functions in the plane and thus $\Ket{n_r, l, n_z}$ orbitals.
This strictly unidimensional system with periodic boundary conditions can then provide a permanent lattice for the aforementioned fundamental physics mesoscale ring experiments such as the Aharanov-Bohm effect and persistent currents.

\section{Conclusion}

\paragraph*{} We demonstrated user-friendly one-pot synthesis of both \SI{3}{} and \SI{4}{ML} CdS/CdSe core/crown heterostructures.
The synthesis method yields better defined structures than continuous injection method.
In these heterostructures, excitons generated in the CdS NPLs are efficiently transferred and confined to the CdSe outer ring.
Ring width is controlled by the amount of Se precursor added to the synthesis.
At respectively low and high amount of Se, QD-like CdSe and CdSe NPL photoluminescence spectra are observed.
This shows a change in the electronic dimensionality from 0D to 2D with an intermediate 1D dimensionality ring.
The emission maximum can be adjusted from \SI{2.42}{eV} to \SI{2.90}{eV} (\SI{428}{nm} to \SI{512}{nm}) using both the CdS NPL seed thickness and the CdSe ring width, giving two degrees of freedom to the tuning of the optical properties.
Hence, we part with the rigid electronic structure of standard NPLs and are now able to continuously tune the electronic and optical properties of a CdS/CdSe lateral heterostructure.

\paragraph*{} The non-injection synthetic method should be applicable to the growth of CdSe around other NPLs.
Also, the quantum ring geometry grants electronic structure fine-tuning to NPLs that, despite their advantages over cQDs, could only be discretely tuned. As such, this geometry could advantageously be applied to other II-VI semiconductors than CdSe. Furthermore, quantum rings could be synthesized with semiconductor alloys, which would lead to an even greater control over their band alignment.
While this article identified the 0D and 2D geometries in photoluminescence and absorption spectra, the 1D quantum ring state was only interpolated. 
An extensive study of these spectra with fine variation of Se amount around the observed shift from 0D to 2D behavior is needed to systematically synthesize true 1D heterostructure.

\begin{acknowledgement}
The research was supported by National Science and Engineering Research Council(NSERC). The authors thank Patrick Larochelle for technical support, Denis Boudreau for sharing experimental equipment, and Dominic Larivière as well as Anna Ritcey for their inputs on the manuscript and interesting discussions.

\end{acknowledgement}

\paragraph*{Supplementary materials:}Five figures: TEM micrographs for \SI{3}{ML} sample, EDS analysis example, X-ray diffractogram, PLE spectra for each sample, and relation between PL$_\text{max}$ and FWHM.

\bibliography{cQR_Article}

\end{document}